\PassOptionsToPackage{unicode}{hyperref}
\PassOptionsToPackage{hyphens}{url}
\PassOptionsToPackage{dvipsnames,svgnames,x11names}{xcolor}
\documentclass[
  12pt]{article}

\usepackage{amsmath,amssymb}
\usepackage{soul}
\usepackage{iftex}
\ifPDFTeX
  \usepackage[T1]{fontenc}
  \usepackage[utf8]{inputenc}
  \usepackage{textcomp} % provide euro and other symbols
\else % if luatex or xetex
  \usepackage{unicode-math}
  \defaultfontfeatures{Scale=MatchLowercase}
  \defaultfontfeatures[\rmfamily]{Ligatures=TeX,Scale=1}
\fi
\usepackage{lmodern}
\ifPDFTeX\else  
\fi
\IfFileExists{upquote.sty}{\usepackage{upquote}}{}
\IfFileExists{microtype.sty}{% use microtype if available
  \usepackage[]{microtype}
  \UseMicrotypeSet[protrusion]{basicmath} % disable protrusion for tt fonts
}{}
\makeatletter
\@ifundefined{KOMAClassName}{% if non-KOMA class
  \IfFileExists{parskip.sty}{%
    \usepackage{parskip}
  }{% else
    \setlength{\parindent}{0pt}
    \setlength{\parskip}{6pt plus 2pt minus 1pt}}
}{% if KOMA class
  \KOMAoptions{parskip=half}}
\makeatother
\usepackage{xcolor}
\makeatletter
\ifx\paragraph\undefined\else
  \let\oldparagraph\paragraph
  \renewcommand{\paragraph}{
    \@ifstar
      \xxxParagraphStar
      \xxxParagraphNoStar
  }
  \newcommand{\xxxParagraphStar}[1]{\oldparagraph*{#1}\mbox{}}
  \newcommand{\xxxParagraphNoStar}[1]{\oldparagraph{#1}\mbox{}}
\fi
\ifx\subparagraph\undefined\else
  \let\oldsubparagraph\subparagraph
  \renewcommand{\subparagraph}{
    \@ifstar
      \xxxSubParagraphStar
      \xxxSubParagraphNoStar
  }
  \newcommand{\xxxSubParagraphStar}[1]{\oldsubparagraph*{#1}\mbox{}}
  \newcommand{\xxxSubParagraphNoStar}[1]{\oldsubparagraph{#1}\mbox{}}
\fi
\makeatother

\usepackage{longtable,booktabs,array}
\usepackage{calc} % for calculating minipage widths
\usepackage{etoolbox}
\makeatletter
\patchcmd\longtable{\par}{\if@noskipsec\mbox{}\fi\par}{}{}
\makeatother
\IfFileExists{footnotehyper.sty}{\usepackage{footnotehyper}}{\usepackage{footnote}}
\makesavenoteenv{longtable}
\usepackage{graphicx}
\makeatletter
\def\maxwidth{\ifdim\Gin@nat@width>\linewidth\linewidth\else\Gin@nat@width\fi}
\def\maxheight{\ifdim\Gin@nat@height>\textheight\textheight\else\Gin@nat@height\fi}
\makeatother
\setkeys{Gin}{width=\maxwidth,height=\maxheight,keepaspectratio}
\makeatletter
\def\fps@figure{htbp}
\makeatother

\makeatletter
\@ifpackageloaded{caption}{}{\usepackage{caption}}
\AtBeginDocument{%
\ifdefined\contentsname
  \renewcommand*\contentsname{Table of contents}
\else
  \newcommand\contentsname{Table of contents}
\fi
\ifdefined\listfigurename
  \renewcommand*\listfigurename{List of Figures}
\else
  \newcommand\listfigurename{List of Figures}
\fi
\ifdefined\listtablename
  \renewcommand*\listtablename{List of Tables}
\else
  \newcommand\listtablename{List of Tables}
\fi
\ifdefined\figurename
  \renewcommand*\figurename{Figure}
\else
  \newcommand\figurename{Figure}
\fi
\ifdefined\tablename
  \renewcommand*\tablename{Table}
\else
  \newcommand\tablename{Table}
\fi
}
\@ifpackageloaded{float}{}{\usepackage{float}}
\floatstyle{ruled}
\@ifundefined{c@chapter}{\newfloat{codelisting}{h}{lop}}{\newfloat{codelisting}{h}{lop}[chapter]}
\floatname{codelisting}{Listing}

\makeatother
\makeatletter
\@ifpackageloaded{caption}{}{\usepackage{caption}}
\@ifpackageloaded{subcaption}{}{\usepackage{subcaption}}
\makeatother

\ifLuaTeX
  \usepackage{selnolig}  % disable illegal ligatures
\fi
\usepackage[]{natbib}
\usepackage{bookmark}

\IfFileExists{xurl.sty}{\usepackage{xurl}}{} % add URL line breaks if available
\hypersetup{
  pdftitle={Title},
  pdfauthor={Author 1; Author 2},
  pdfkeywords={3 to 6 keywords, that do not appear in the title},
  colorlinks=true,
  linkcolor={blue},
  filecolor={Maroon},
  citecolor={Blue},
  urlcolor={Blue},
  pdfcreator={LaTeX via pandoc}}

\newcommand{\anon}{1}

\begin{document}

\def\spacingset#1{\renewcommand{\baselinestretch}%
{#1}\small\normalsize} \spacingset{1}

%%%%%%%%%%%%%%%%%%%%%%%%%%%%%%%%%%%%%%%%%%%%%%%%%%%%%%%%%%%%%%%%%%%%%%%%%%%%%%

\if1\anon
{
\title{\bf A Statistical Framework for Signal Extraction in Noisy Ionospheric Observations from the International Space Station}

\author{%
  \begin{tabular}{c}
    Rachel Ulrich$^{1}$, Kelly R. Moran$^{1}$, Ky Potter$^{1,2}$, \\
    Lauren A. Castro$^{3}$, Gabriel R. Wilson$^{4}$, Carlos Maldonado$^{4}$
  \end{tabular}
  \\[0.9em]
  \small
  \begin{tabular}{@{}r p{0.78\textwidth}@{}}
    $^{1}$ & Statistics, Computing and Artificial Intelligence Division, Los Alamos National Laboratory \\
    $^{2}$ & Statistics and Actuarial Science Department, Simon Fraser University \\
    $^{3}$ & Analytics, Intelligence and Technology Division, Los Alamos National Laboratory \\
    $^{4}$ & Space Science and Applications, Intelligence and Space Research Division, Los Alamos National Laboratory
  \end{tabular}
}
\maketitle
} \fi

\bigskip
\begin{abstract}
The Electric Propulsion Electrostatic Analyzer Experiment (ÈPÈE) is a compact ion energy bandpass filter deployed on the International Space Station (ISS) in March 2023 and providing continuous measurements through April 2024. This period coincides with the Solar Cycle 25 maximum, capturing unique observations of solar activity extremes in the mid- to low-latitude regions of the topside ionosphere.  From these in situ spectra we derive plasma parameters that inform space-weather impacts on satellite navigation and radio communication. We present a statistical processing pipeline for ÈPÈE that (i) estimates the background current profile of the instrument, (ii) accounts for irregular temporal sampling, and (iii) extracts ionospheric signals. Rather than discarding data below some current threshold, the method learns an instrumental baseline and fits the residual measurement surface using a scaled Vecchia Gaussian process approximation, decreasing the number of observation times typically rejected by thresholding by over 98\%. The resulting products increase data coverage and enable noise-assisted monitoring of ionospheric variability.
\end{abstract}

\noindent%
{\it Keywords:} Gaussian processes, Scaled Vecchia, Ionospheric science, signal processing

%\vfill
%\newpage
\spacingset{1.8} % DON'T change the spacing!

\section{Introduction}\label{sec-intro}

\subsection{Background on EPEE instrument}
\label{background_EPEE_instrument}
The Electric Propulsion Electrostatic Analyzer Experiment (ÈPÈE) is a low-cost, rugged, compact laminated electrostatic analyzer (ESA) deployed on the International Space Station (ISS) in March 2023. ÈPÈE provided continuous measurements from its deployment through April 2024, spanning the entry into the maximum of Solar Cycle 25 \citep{Maldonado2025}. In space plasma physics, electrostatic analyzers (ESAs) are widely used for measuring particle energy per unit charge ($E/q$). By combining sensor geometry with an applied electric field to form an energy band-pass filter, ÈPÈE measures ion energy and current from local charged particle populations \citep{Maldonado2025, Maldonado2023}. From these measurements, macroscopic parameters of interest can be derived. 

The ISS orbits at  $\sim400$ kilometers above the Earth's surface, providing critical measurements of the top-side ionosphere directly above the F2 peak that cannot be obtained from ground-based methods \citep{Maldonado2025}. The ionosphere is a low-density, "cold" plasma dictating that currents are relatively small (reported in nanoamperes) and temperatures are low (0.1-0.2 electron Volts). Accurate plasma measurements facilitate understanding of the complex relationships between solar storms, the solar cycle, eclipse, and spacecraft charging. This understanding is essential not only for exploring phenomena of scientific interest (e.g., the Equatorial Ionization Anomaly and Traveling Ionospheric Disturbances), but also for anticipating space-weather effects on the growing number of low Earth orbit satellites that support navigation, communication, and national security \citep{minow_iswat_2024}. 

\begin{figure}
    \centering
    \includegraphics[width=\textwidth]{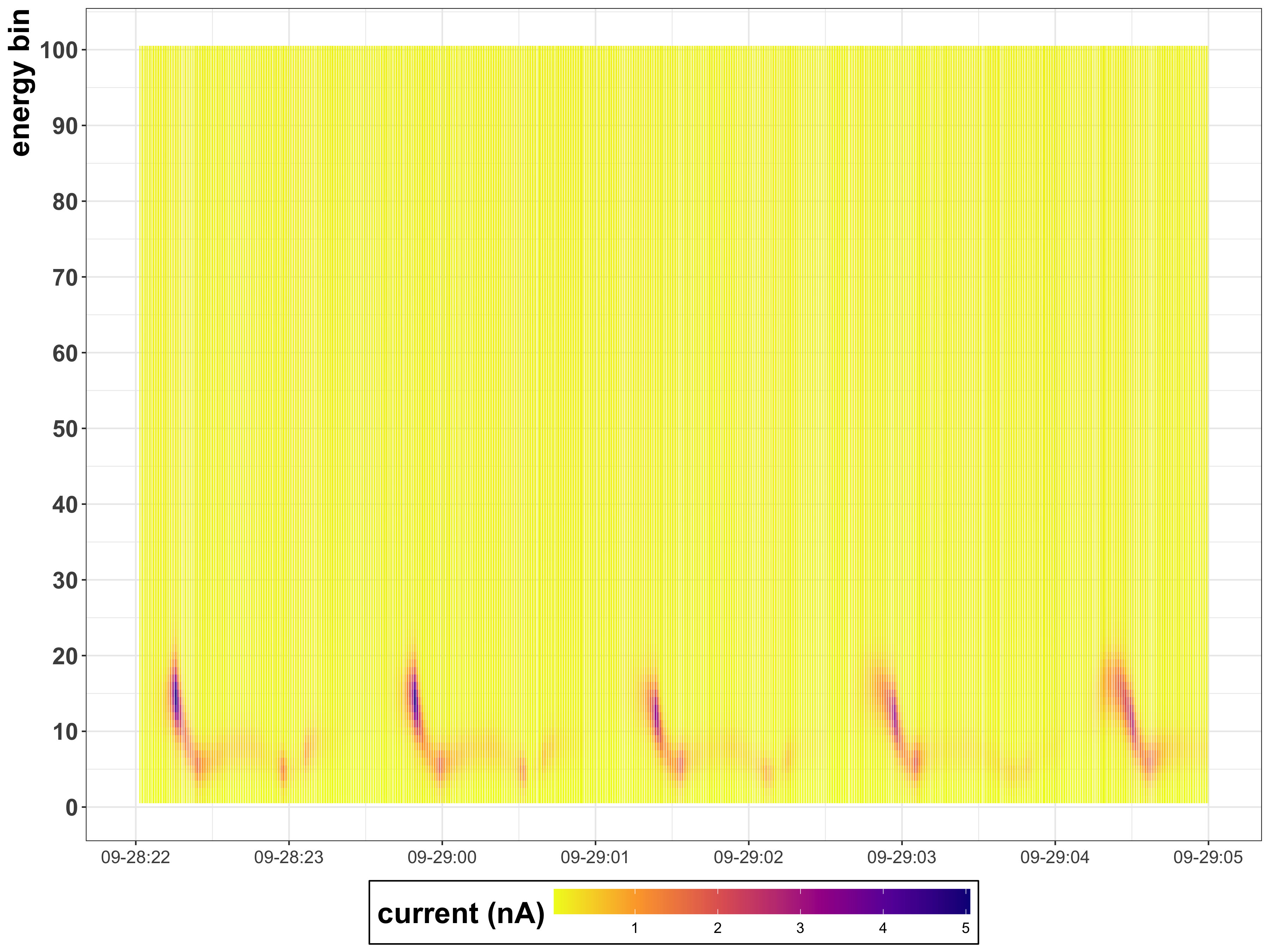}
    \caption{Heat map of current value by energy bin for the time range of interest (Sept 28, 2023 22:00:00 UTC - Sept 29, 2023 05:00:00 UTC). Color indicates the range of current values between 0-5 nA across the 100 discrete energy bins, displayed on the y-axis. Time is on the x-axis (MM-dd:HH). For this time period, larger current values mainly occur in energy bins at or below bin 25, with the majority of current values recorded near the noise floor (0.15 nA) in higher energy bins.}
    \label{fig:fig_image}
\end{figure}

The ÈPÈE instrument provides measurements at a rate of 0.5 Hz, resulting in a single data point approximately every two seconds. Due to the selective filter design, at each timestamp ($t$) a measurement of the current ($\mathbf{I}$) and energy ($\mathbf{E}$) in electron Volts (eV) is recorded for each of 100 discrete energy bins ($\mathbf{B}$ with $b=1,2,...,B$). Energy values are discretized into bins with the total observed range spanning approximately 0.8 - 185 eV. At each timestamp we have values of current across energy bins ($I_{t,b}$), or a distribution of current that characterizes the local space plasma at that moment in time. Figure \ref{fig:fig_image} displays a heat map of current values across energy bins for the seven-hour time period that serves as the example for this paper (Sept 28, 2023 22:00:00 UTC - Sept 29, 2023 05:00:00 UTC). The majority of higher current values (ranging from 0-5 nA) occur at or below energy bin 25, with current values recorded near the instrument noise floor (0.15 nA) for higher energy bins. This enhancement pattern recurs approximately every  $\sim90$ minutes, consistent with the ISS orbital period. 

Interpretation of ÈPÈE's measurements relies on basic principles of plasma physics. Recalling that a plasma consists of electrons, ions, and neutrals, each species can be described by a distribution function $f_\alpha(\mathbf{r},\mathbf{v},t)$
that evolves under external forces and collisions, where $\mathbf{r}$ and $\mathbf{v}$ are the position and velocity vectors and $t$ is time. This function, often referred to as the \textit{phase space density}, represents the number of particles per unit volume in real and velocity space — that is, the particle density at a given velocity, position, and time. Because individual motions cannot be observed, macroscopic parameters—such as density and temperature—are estimated by taking moments of these distributions \citep{Howard2002}.

Due to ÈPÈE’s limited field of view, the movement of space plasma relative to the detector face is primarily along the velocity vector, allowing the use of a one-dimensional Maxwellian distribution, 
$f(v_x)$, as an analytical approximation to the ion velocity distribution. Here, $f(v_x)$ is a species-specific simplification of the general distribution $f_\alpha(\mathbf{r},\mathbf{v},t)$, assuming spatial and temporal homogeneity and that perpendicular velocity components contribute negligibly within ÈPÈE’s narrow field of view. However, because the plasma cannot be considered to be in thermal equilibrium owing to external interference from the ISS, a \textit{drifted Maxwellian} is used to account for the resulting bulk flow velocity. This distribution depends on 
$n,m,k,T$ and the drift velocity $v_{x,d}$, where $k$ is the Boltzmann constant, $m,T$ are species-specific (with $T$ representing the thermal spread in local equilibrium), and $n$ is the total number density serving as the normalizing constant \citep{Maldonado2023}:

\begin{equation}
    f(v_x) = n(\frac{m}{2\pi kT})^{1/2} \ \exp(\frac{-m}{2kT}(v_x-v_{x,d})^2)
    \label{eq:maxwellian_dist}
\end{equation}

The form of this distribution is familiar, as it is a special application of the Gaussian distribution. 

ÈPÈE measures ion current arising from the flux of charged particles transmitted through its energy bandpass at each voltage step. The measured current 
\( I \propto q \!\! \int v_x f(v_x)\,dv_x \) therefore reflects the velocity distribution weighted by particle speed within the analyzer’s selected energy-per-charge window. The current peak $( I_{0,t} )$ occurs at the most probable energy $( E_{0,t} )$  linking the drifted Maxwellian form of Eq. \ref{eq:maxwellian_dist} to the subsequent energy relation used to infer the spacecraft potential 
\( \phi_t \) \citep{Maldonado2023}:

\begin{equation}
    E_{0,t} = \frac{1}{2}mv_x^2 + q\phi_{t}
        \label{eq:sc_potential_equation}
\end{equation}

Thus, selecting the maximum current value ($I_{0,t}$) and respective energy value ($E_{0,t}$) at a given timestamp allows us to solve for spacecraft potential ($\phi_{t}$). Recall $q$ is the particle charge (see Section \ref{background_EPEE_instrument}). 

\begin{figure}
    \centering
    \includegraphics[width=\textwidth]{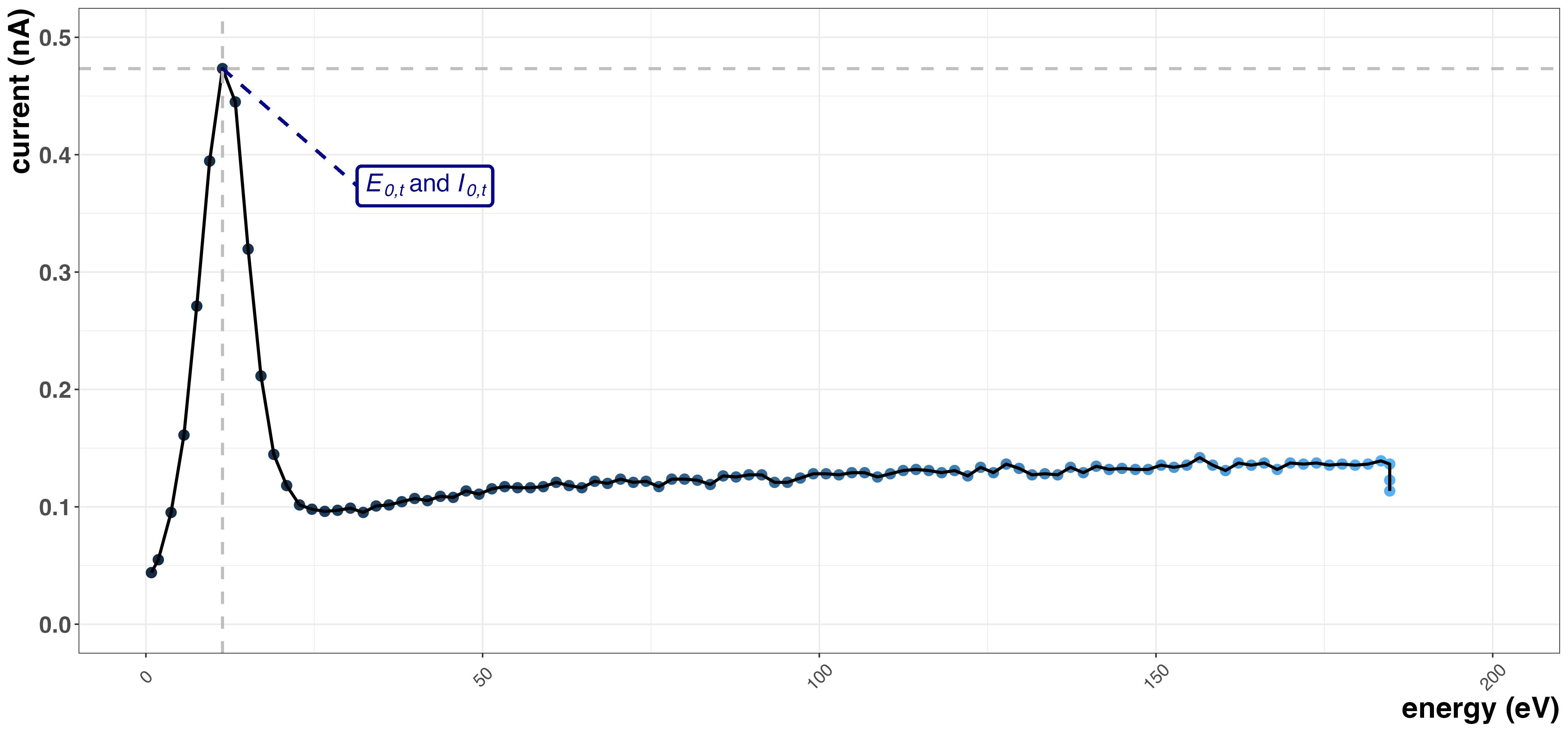}
    \caption{Distribution of current values across energy bins for a single timestamp (2023-09-29 00:33:18 UTC). Discrete current observations are marked by points and the underlying empirical distribution is fit with a line. Energy bin is on the x-axis and current is on the y-axis (nA) with color referring to energy value (eV). Energy value is discretized by bins; energy values increase as bin number increases. This is referenced with indigo blues fading to azure as energy value increases across the x-axis. Gray lines indicate the intersection of the maximum values for current and energy.}
    \label{fig:E0_I0_Maxwellian_figure}
\end{figure}

Figure~\ref{fig:E0_I0_Maxwellian_figure} demonstrates the distribution of current values for a single moment in time under the conditions in which a Maxwellian is a reasonable approximation. The peak current value describes the most likely value and is matched to the energy value for the respective energy bin.

\subsection{Background on Spacecraft Charging}
\label{sec:background_spacecraft_charging}

The electrical charging physics of the ISS are complex. In brief, the design and motion of the ISS and interaction with the ionosphere produce conditions potentially hazardous to astronauts during extravehicular activity, and to the surface of the ISS itself (please see \citet{Hastings1995} for a detailed explanation).  Initially, the Floating Potential Measurement Unit (FPMU) was deployed in 2006 as a permanent diagnostic tool of charging physics, designed with four separate instruments that together produced reliable measurements of ambient plasma density and electron temperature in addition to frame potential. These observations have redundant measurements over several instruments: Density is directly measured by the Wide-sweep Langmuir Probe (WLP) and the Narrow-Sweep Langmuir Probe (NLP), and derived from measurements taken by the Plasma Impedance Probe (PIP) while spacecraft charge is measured by the WLP and NLP, and referenced by the FPP \citep{Swenson2003, Minow2023}. These redundancies facilitate cross-validation between instruments, rendering measurements from the FPMU to be the general source of truth by the scientific community. 

Although intended for only three years operation, the FPMU has provided observations of local space plasma for nearly two decades \citep{Minow2023}. However, these data were not a continuous record, rendering EPEE the main provider of data for its limited lifetime deployment. A seven-hour period of overlap exists between the two data sets in 2023, on Julian days 271 and 272. This period of overlap will be used as the case example - and opportunity for comparison - for the remainder of the analysis in instances where subsets of the data record are most efficacious for visualization.

\begin{figure}
    \centering
    \includegraphics[width=\textwidth]{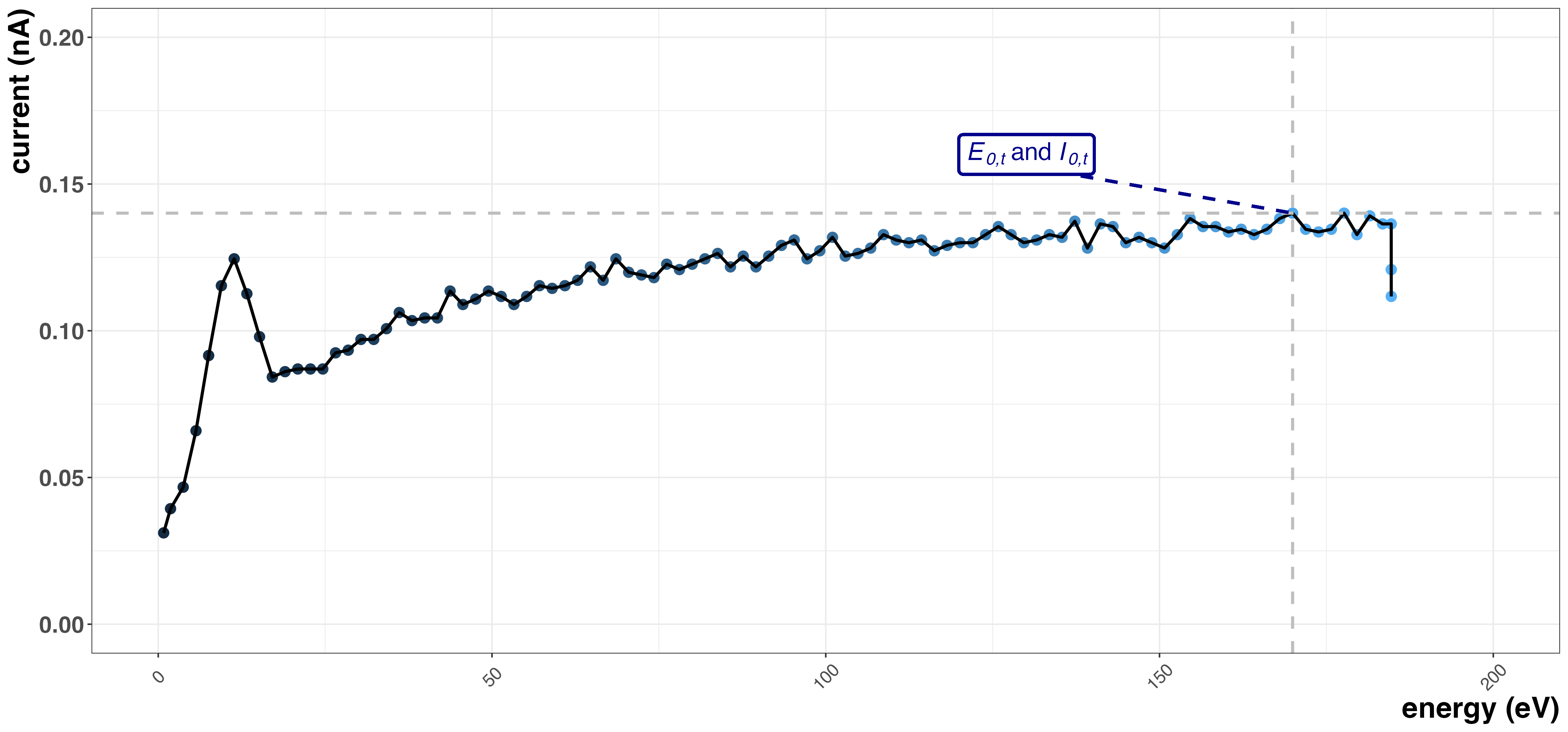}
    \caption{Empirical distribution of current values across energy bins for a single timestamp (2023-09-28 23:05:20 UTC). Energy bin is on the x-axis and current is on the y-axis (nA). The increase in energy value (eV) is represented through a lightening blue color palette, with indigo blues indicating lower energy values and azure blues indicating higher energy values. In this instance, choosing a maximum value is not clear, and the majority of observations fall in close proximity to the actual maximum.}
    \label{fig:E0_I0_empirical_figure_day271_chunk4}
\end{figure}

As shown in Equation ~\eqref{eq:sc_potential_equation}, determining spacecraft charge at a given timestamp requires the selection of the maximum current value and respective energy value, resulting in data formatted with a single current value, energy bin, energy value and spacecraft charge value per timestamp. If, however, our assumed underlying distribution does not reasonably approximate the data, taking the maximum current value does not align with our expectation. Below, Figure~\ref{fig:E0_I0_empirical_figure_day271_chunk4} shows an empirical distribution of current values for a single timestamp with no clear maximum. 

Figure~\ref{fig:E0_I0_empirical_figure_day271_chunk4} demonstrates another, related issue. Although not always the case, the timestamps that present difficulties in properly selecting the maximum value are often hovering at or below the instrument noise floor (0.15 nA). Observations near the instrument's sensitivity threshold are notoriously difficult to handle, as the true signal can be hard to extract. Moreover, maximum current values around this threshold often occur in tandem with inflated energy values. As energy has a direct relationship with spacecraft potential (see Equation\eqref{eq:sc_potential_equation}), very high energy values result in clearly erroneous spacecraft potential values, far above the automatic discharge threshold of the ISS (spacecraft potential of 40 volts) \citep{Swenson2003}. Initial thresholding efforts considered these points ``untrustworthy'' and removed them from subsequent analyses, creating a notable amount of missing data, particularly during the 7-hour overlap period used in cross-calibration efforts with FPMU. 

\begin{figure}[ht]
    \centering
    \includegraphics[width=\textwidth]{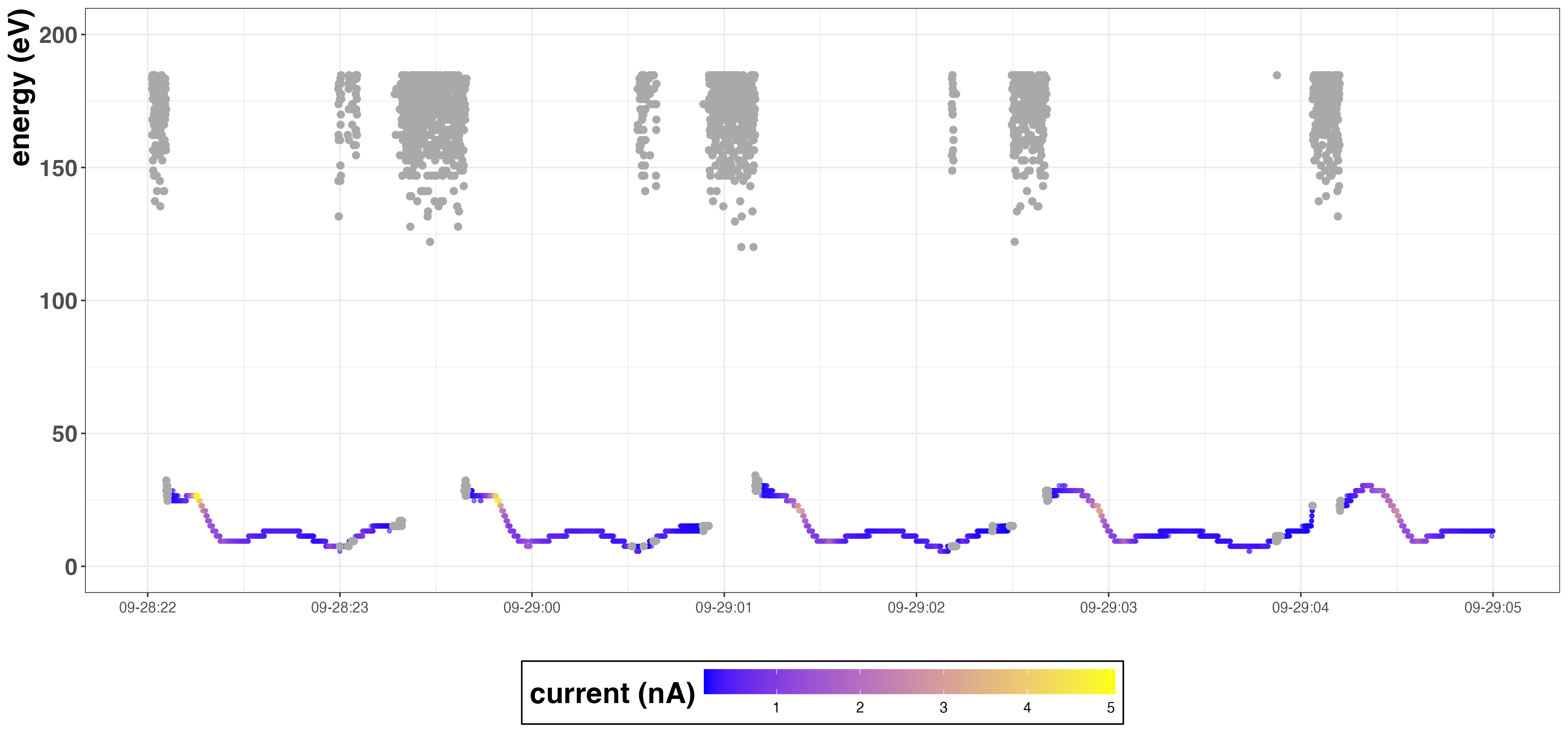}
     \caption{Maximum current values (y-axis) across time (x-axis), with colors indicating respective current values. Cooler colors indicate lower current values and warmer colors indicate higher current values (range 0-5 nA). All observations with current values at or below the instrument noise floor (0.15 nA) are colored in dark grey. These current values often - although not always - occur in tandem with very high energy values.}
    \label{fig:FPMUtime_rawEPEE_nAeV_noise}
\end{figure}

Figure \ref{fig:FPMUtime_rawEPEE_nAeV_noise} shows the overlap period for Julian days 271-272 in 2023 for EPEE maximum current ($I_{0,t}$) and energy ($E_{0,t}$) values. Energy values are on the y-axis with time on the x-axis (month-day:hour) and colors pertaining to current values. All current values at or below the instrument noise floor (0.15 nA) are colored in dark grey, and the remaining current values (ranging from 0.15-5 nA) are represented with darker blues indicating lower values and yellows indicating higher values. Current values at or below the noise floor predominantly occur in tandem with very high energy values, as exemplified in Figure \ref{fig:E0_I0_empirical_figure_day271_chunk4}. Note that the initial respective energy value for the raw current maximum would be 169.9 eV, a clearly unphysical value. 

Furthermore, when these data are aligned with sensors providing spacecraft potential observations from the FPMU instrument (Wide-Sweeping Langmuir Probe (WLP), Floating Potential Probe (FPP); please see \citet{Swenson2003} for more information on these sensors), more instances of missingness occur. Due to mismatching time stamps and missing data from FPMU, the already short window of time available for cross-calibration becomes further disjointed, limiting our ability to extract as much information as possible from the data. Typically, these values would be dropped from analysis but even seemingly erroneous observations may provide motivation for future explorations, thus we seek to preserve as many observations as possible. 

For the remainder of the paper we consider an illustrative subset of data that overlaps with FPMU availability (Sept 28, 2023 22:00:00 UTC - Sept 29, 2023 05:00:00 UTC) to explore an alternative approach to handling noisy current input values, employing a series of techniques to capture data points previously considered unusable and smooth input values for use in the calculation of spacecraft potential. We show how this approach facilitates better cross-calibration with FPMU data and estimation of parameters crucial to mission safety and success.

\section{Methods}\label{sec-meth}

\subsection{Overview}

As previously described, the instrument noise floor hovers around 0.15 nA, leading to mis-identification of the appropriate peak energy bin. An alternative approach to handling these values is outlined in the following steps:

\begin{enumerate}
    \item Learn the smooth current surface: Fit Gaussian process (GP) models to the current data $\mathbf{I}$, producing smooth estimates $\mathbf{I^S}$ of the current surface as a function of time and energy.
    
	\item Identify ``minimal signal'' timestamps:  Using $\mathbf{I}$, identify a subset of times that are background-dominated $\mathbf{t}^f=\{t_{k_1}^f,t_{k_2}^f,…,t_{k_M}^f\}$ with $ k_1,k_2,.., k_M$ indexing the selected timestamps.

	\item Iteratively fit and refine an instrument baseline profile, i.e. the shape around which pure-noise observations at each energy bin appear: Consider the smooth current profile $I^S_{k_m}(b)$ associated with time $t_{k_m}$ across energy bins $b \in \{1,\ldots,100\}$. We model this profile as the sum of two main components: a Richards curve with a parabolic adjustment, which is our baseline profile $N^*_{k_m}(b)$ or ``background'', plus a Gaussian impulse to capture ``signal''. We do this for each profile, then select from these individual profiles to identify a conservative underlying baseline profile $N^*(b)$.
    
	\item Subtract the baseline profile, post-process to remove any remaining noisy artifacts, and calculate downstream metrics: Let $I^*(t,b)=I^S(t,b)-N^*(b)$ denote the set of all smoothed, baseline-subtracted, cleaned current values. Using these $I^*$ values in place of $I$, we can compute the maximum current value at timestamp $t$ as $I_{(0,t)}^*$ in order to retain more usable observations.
 
\end{enumerate}

\begin{figure}
    \centering
    \includegraphics[width=\textwidth]{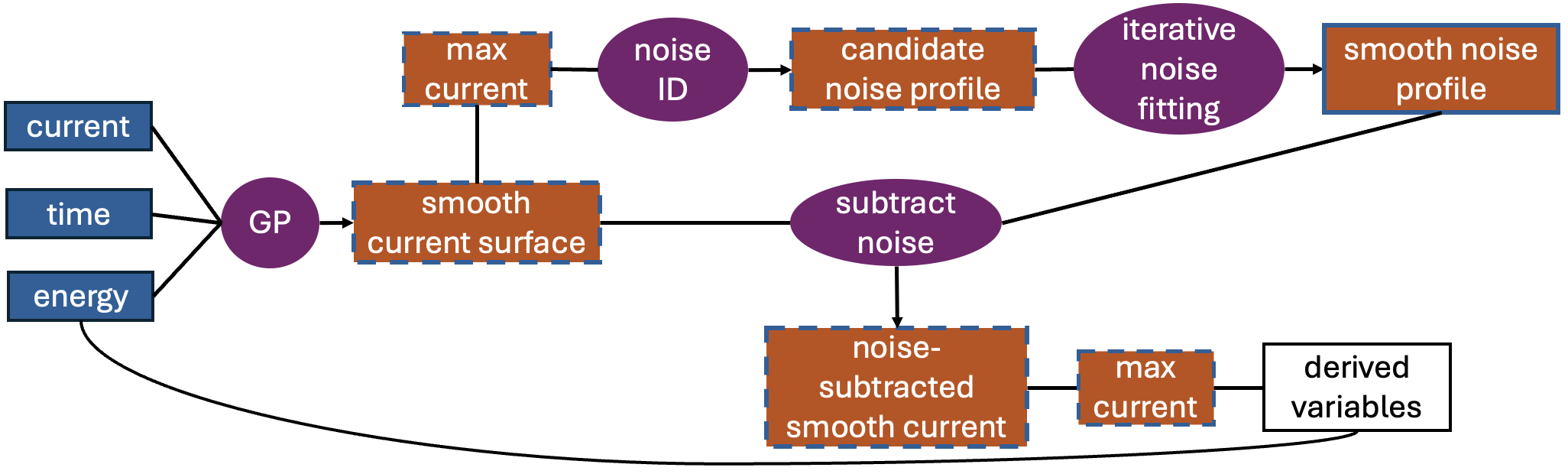}
    \caption{Flowchart depicting the methodology for handling values near the instrument baseline. Inputs are in blue and outputs in burnt orange, both are rectangular shapes. Outputs that then become inputs in subsequent steps have a dashed outline. Processes are in purple circles with a solid line representing input to a process and an arrow representing output of a process. The final, desired variables of interest are identified with a white box outlined in black.}
    \label{fig:process_flowchart}
\end{figure}

Figure \ref{fig:process_flowchart} shows a different perspective of these steps, with inputs (teal blue rectangles), outputs (burnt orange rectangles) and processes (plum circles) showing movement from observation-based input to the derived variables of interest (white box with black outline). 

Although  specific decisions have been made at various steps that were best-suited to these data and application, the overall methodology represents a general denoising approach. For example, in Step 1 we implement a Vecchia GP in order to learn the smooth current surface; a different underlying model could be chosen, a different GP approximation method could replace scaled Vecchia, or - depending on the size of the data - an exact GP could be implemented. The baseline itself could be described with a known shape (e.g. a constant offset), if this were appropriate, instead of learning this profile from the data. 

\subsection{Step 1: Fit a smooth surface using a GP approximation}

The data are partitioned into quarter-day intervals and a GP is used to learn the underlying smooth current surface as a function of time and energy for each 6-hour chunk.
GPs are flexible statistical models that allow for surfaces to be fit to data without requiring many assumptions to be made about the surface structure. GPs can be conceptualized as distributions over functions (or surfaces) or as infinite-dimensional generalizations of the multivariate Gaussian distribution.  GPs are fully specified by two functions, a mean function ($m: \mathbf{X}\rightarrow \mathbb{R}$) and a covariance function ($k: \mathbf{X}\times \mathbf{X} \rightarrow \mathbb{R}$). The mean function is commonly set to zero for centered data, leaving the dependence structure between observations to drive the model’s behavior. The current surface over time and energy is then defined:

\begin{equation}
    I{(x_i)} = I^S(x_i) + \epsilon_i, \quad I^S(\cdot) \sim \mathcal{GP}(0,k(\cdot,\cdot)), \quad \epsilon_i\sim N(0, \sigma^2)
\end{equation}

Let $x_i=(t_i,e_i)$ denote the time/energy combination associated with observation $i$. Then, the vector $\mathbf{I} = (I(x_1),..., I(x_n))^T$ of observed currents at inputs $x_1,\ldots,x_n$ follows an $n$-variate Gaussian distribution with covariance matrix $K + \sigma^2 I_n$, where the $(i,j)$th entry of $K$ is given by $K_{ij} = k(x_i,x_j)$. Thus, the kernel $k(\cdot,\cdot)$ specifies the covariance between latent current surface values at two time--energy combinations \citep{Betancourt2020, Lawrence2022}. We define $k(\cdot, \cdot)$ as a Mat\'ern $3/2$ kernel \citep{Rasmussen2006} with separate range parameters for time and energy (see Appendix~\ref{app:GP_vecchia_details}). These range parameters, the process variance, and the noise variance $\sigma^2$ are optimized via maximum likelihood.

Inverting the covariance matrix in the likelihood computation becomes infeasible as $n$ increases due to the required $O(n^3)$ operations. Many approaches have been developed to circumvent this issue. One such approximation, the Vecchia approximation \citep{Vecchia1988}, was originally developed in spatial statistics and has been substantially extended in recent work. Following the methodology of \citet{Lawrence2022}, and for a fixed ordering of the inputs $x_1,\dots,x_n$, the joint density of the latent current surface evaluated at those inputs is approximated as

\begin{equation}
p\big(I^S(x_1), \dots, I^S(x_n)\big)
\approx
\prod_{i=1}^{n}
p\big(I^S(x_i) \mid I^S(x_{S(i)})\big),
\label{eq:Vecchapprox}
\end{equation}

where $S(i) \subset \{1, \dots, i-1\}$ denotes a fixed-size neighbor set for observation $i$. Details regarding Vecchia design choices are provided in Appendix~\ref{app:GP_vecchia_details}. Because the underlying process is Gaussian, any conditional distribution of the observations is Gaussian; thus each term in \eqref{eq:Vecchapprox} is Gaussian with mean and variance determined by the corresponding submatrices of the Matérn covariance kernel.

\subsection{Step 2: Identify background regions and create a baseline profile}

To identify timestamps that are background-dominated (i.e., those whose global maxima occur away from a local peak, see Figure \ref{fig:E0_I0_empirical_figure_day271_chunk4}), we leveraged the understanding that maxima identified in energy bins above a certain threshold result in non-physical spacecraft potential values (see Section~\ref{sec:background_spacecraft_charging}).
Conservatively, maximum currents identified in bins 50--100 are treated as likely background-dominated maxima. After flagging likely background-dominated timestamps, the next step is to isolate those that represent lowest-signal conditions. To achieve this, the algorithm leverages the clustering of likely background-dominated times, visually evident in Figure \ref{fig:svecch_raw_271_272_comparison} as dense groups of observations at higher energy values. This heuristic approach focuses on identifying the "heart" of these observations, targeting the central regions where background dominance is most pronounced. For these candidate noise timestamps, we computed the total decrease in current across the first 20 energy bins, which is expected to be zero for a background-only profile. To favor timestamps lying in the center of sustained background-dominated periods rather than at their edges, we applied a centered rolling window to this quantity over the ordered candidate background timestamps. Starting with a window spanning 3 timestamps and increasing it in odd-sized increments (3, 5, 7, \dots), a timestamp was retained only if the rolling sum within its window was zero. The window was expanded until the number of retained timestamps would fall below a prespecified target proportion of candidate background-dominated times (default 10\%), and the retained set from the previous step was used. Larger windows enforce this zero-decrease condition over a broader neighborhood, producing a smaller but more conservative set of background-only timestamps, whereas smaller windows retain more timestamps but are less stringent. This procedure isolates the most stable, minimal-signal intervals and enables accurate characterization of the instrument's baseline background-dominated behavior.

After identifying background timestamps, we model each background-dominated current profile $I^S_{k_m}(b)$ as the sum of an instrument baseline and a possible ionospheric signal contribution via a profile-fitting procedure that models $I^S_{k_m}(b)$ as the sum of three components: a Richards curve, a parabolic curve, and a Gaussian peak.

\begin{equation}
    I^S_{k_m}(b) \approx 
    \underset{\text{instrument baseline}}{\underbrace{f_{k_m, \mathrm{R}}(\boldsymbol{\theta}_R) + f_{k_m, \mathrm{P}}(\boldsymbol{\theta}_P)}} + \underset{\text{ionospheric signal}}{\underbrace{f_{k_m, \mathrm{G}}(\boldsymbol{\theta}_G)}}
\end{equation}

where the instrument baseline is composed of the Richards curve fit ($f_{k_m, \mathrm{R}} (\boldsymbol{\theta}_R)$) and parabolic component ($f_{k_m, \mathrm{P}}(\boldsymbol{\theta}_P)$), parameterized by $\boldsymbol{\theta}_R = (A, k, x_0, \nu, A_0)$ and $\boldsymbol{\theta}_P = (p, q, r)$, respectively.Any remaining localized structure is identified using a Gaussian peak $f_{k_m,\mathrm{G}}(\boldsymbol{\theta}_G)$ with parameters $\boldsymbol{\theta}_G = (\alpha, \mu, \sigma)$ (see Appendix \ref{app:noise-model} for full mathematical forms). 

The Richards curve is a generalization of the logistic function that introduces a shape parameter $\nu$ to allow for asymmetry in the sigmoidal form. Although originally developed for growth modeling in biology, epidemiology, and ecology \citep{richards1959flexible}, it is well suited to modeling the stabilizing behavior of instrument background across energy bins, capturing smooth rises and plateaus with flexibility in slope and curvature. As such, the Richards component is always included as the base model in our background profile estimate.

After subtracting the Richards fit, a Gaussian component is fit to the residuals to capture any remaining localized structure. This structure is presumed to represent true signal rather than background, so the Gaussian is used only during fitting and not added to the final background profile. Once parameter estimates for the Richards and Gaussian components have stabilized, a parabolic term is introduced to model low-order curvature not captured by the sigmoid. Thus our refined background profile for each timestamp becomes $N^*_{k_m}(b) = f_{k_m, R}( \boldsymbol{\theta}_R) + f_{k_m, P}(\boldsymbol{\theta}_P)$ (see Appendix \ref{app:noise-model} for further details and visualizations). 

Although written as an additive model, the three components are not fit simultaneously. Instead, all parameters are estimated by constrained nonlinear least squares in a sequential procedure designed to prevent localized enhancements from biasing the smooth baseline estimate. Specifically, the Richards component is fit first, a Gaussian term is then fit to the residuals over low energy bins to capture localized signal, and once those two components have stabilized, a parabolic term is fit to the remaining residual structure. The Gaussian component is used only to protect the baseline fit from ionospheric signal and is not included in the final background profile estimate. This sequential strategy avoids joint-fit behavior in which the flexible Richards curve can partially absorb localized peaks, thereby inflating the estimated baseline and suppressing true ionospheric signal. The resulting baseline estimate for each candidate background profile is $N^*_{k_m}(b) = f_{k_m,\mathrm{R}}(\boldsymbol{\theta}_R) + f_{k_m,\mathrm{P}}(\boldsymbol{\theta}_P)$. Among all candidate background profiles, we selected as the final background profile estimate the fitted baseline with the smallest integrated magnitude over bins 2 through 20, corresponding to the most stable low-energy background regime.

\begin{figure}
    \centering
    \includegraphics[width=\textwidth]{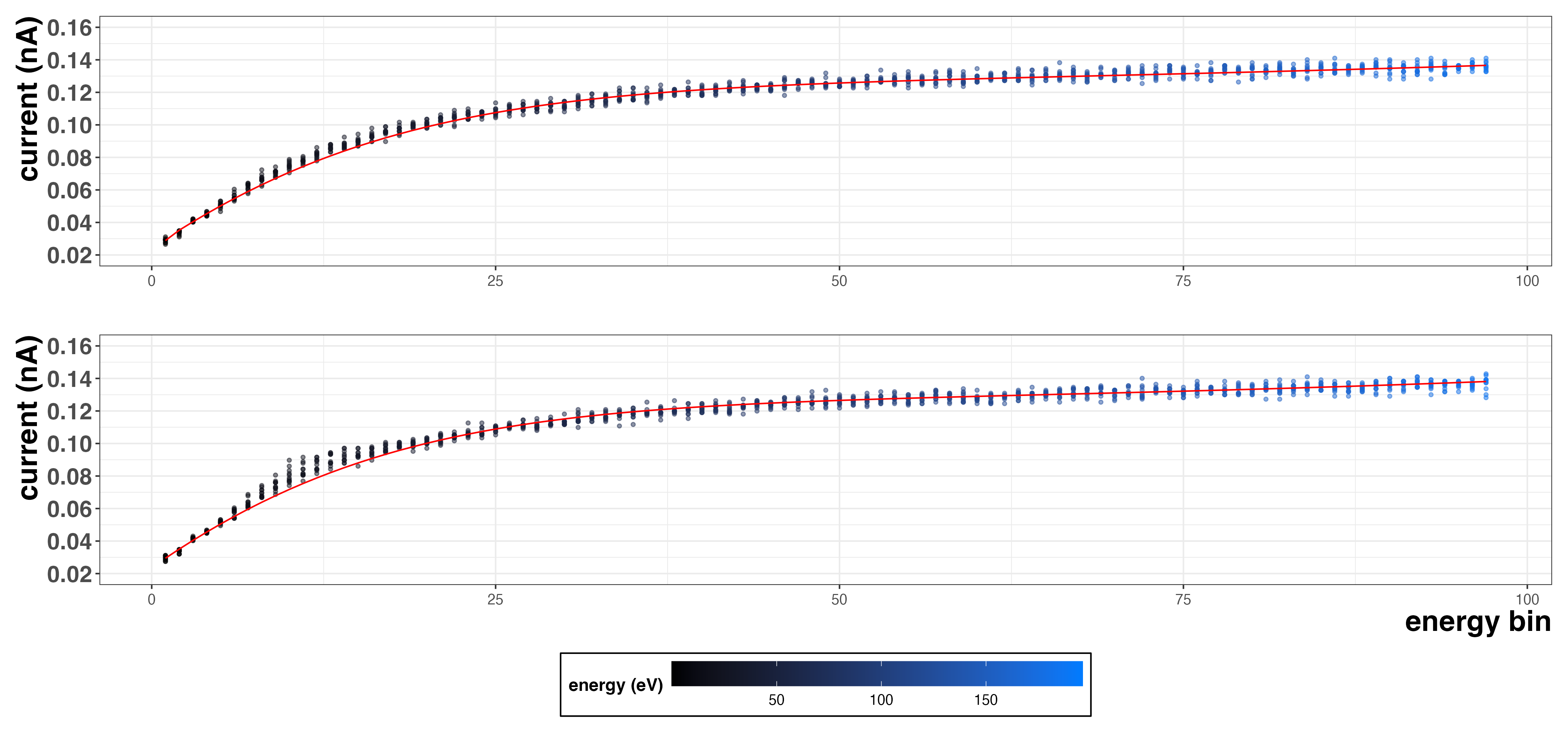}
    \caption{Results of the background profile fitting process for the first (top facet) and fourth quarter (bottom facet) of Julian day 271 (both with n=10799). Candidate background-dominant timestamps (those identified as likely minimal signal) are plotted across energy bins (x-axis) and GP-smoothed current values (y-axis) with raw energy identified by coloring (blues lightening with increasing values). The final fitted background profile, comprised of Richards and parabolic component shapes, is highlighted in red.}
    \label{fig:svecchia_noisefloor_combined}
\end{figure}

Figure \ref{fig:svecchia_noisefloor_combined} shows results for the first and fourth quarters of Julian day 271 (both with n=10799) with GP-smoothed current values (y-axis) for each timestamp plotted across energy bins (x-axis) and color indicating energy value. There are 100 timestamps identified as 'likely minimal signal' from each period. After each timestamp undergoes the fitting process individually, the fitted background profile with the smallest integral is selected (highlighted in red). The first quarter (top facet) candidate selection does a fairly good job and little Gaussian shape is left in the residuals to fit and remove. In contrast, the initial candidate selection in the fourth quarter (bottom facet) picks up signal in the earlier bins, which we successfully capture through the iterative fitting process and preserve by subtracting this shape from the final fitted background profile.

\subsection{Step 3: Learn the smooth current profile}

Because the instrument baseline is omnipresent, we subtract it from our predicted GP fit to yield a learned current surface that is smoother (due to the GP) and less dominated by instrumental effects in low-current times (due to the background identification and subtraction). The background-subtracted, smooth current surface is then used to select the maximum current value by timestamp and the respective energy value from the raw energy values is identified. These maxima represent the inferred peak signal locations and are the only non-constants that enter into the equation for calculating our main downstream metric of interest, spacecraft potential (Equation \eqref{eq:sc_potential_equation}).

\subsection{Step 4: Post-processing}

Although the background-fitting and removal process mitigates many of the previously-deemed untrustworthy times across the 13-month dataset, several post-processing filters are applied to address outlier cases and ensure robust signal identification. Specifically, we remove time intervals that exhibit prolonged sequences of high-energy maxima, which often indicate instrumental anomalies or solar interference. We also exclude observations in which the peak bin undergoes a sudden and drastic downward shift, suggesting an artifact introduced during background profile subtraction, particularly after timeline merging. Finally, any maxima associated with energy values exceeding 45 eV are discarded, as these are considered physically implausible in the context of this analysis. All removed entries are flagged and retained in the final dataset to preserve transparency and support further algorithmic improvements.

\section{Verifications}\label{sec-verify}

\subsection{Background Identification and Instrument Baseline Modeling}

\begin{figure}
    \centering
    \includegraphics[width=\textwidth]{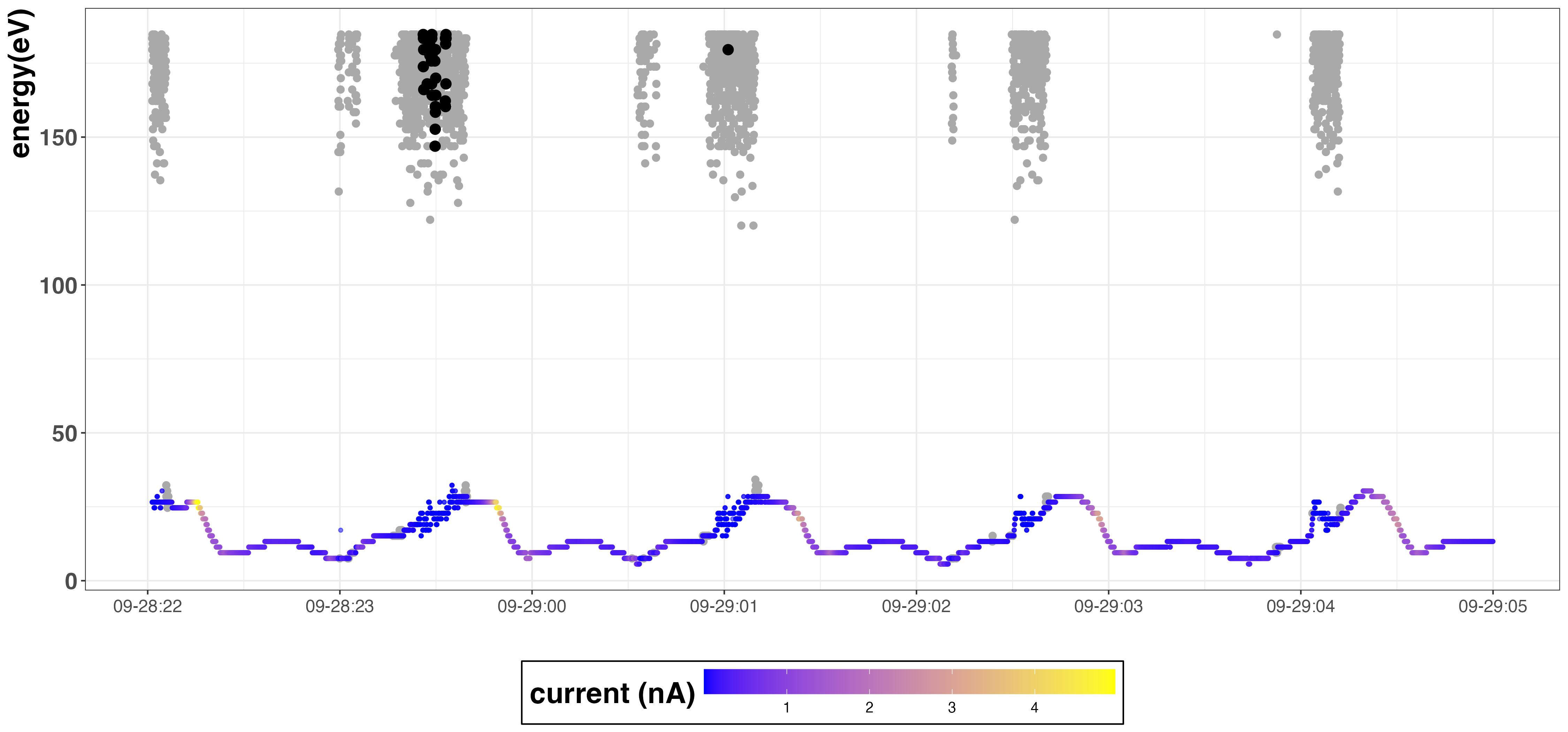}
    \caption{Comparison of original and new methodologies for processing the data. Energies at peak currents found under the new methodology are colored by the baseline-subtracted, smoothed current value with points identified as untrustworthy in black. Original methodology results for observations that were previously dropped from analyses are in grey.}
    \label{fig:svecch_raw_271_272_comparison}
\end{figure}

Figure \ref{fig:svecch_raw_271_272_comparison} shows maximum energy value ($E_0$) results for the 7-hour period spanning Julian days 271-272. Smoothed, noise-subtracted current values are depicted with darker blues indicating lower values and yellows indicating higher values. Time is on the x-axis. Observations identified as untrustworthy and dropped through the original thresholding methodology are colored in gray. The black points represent observations that were statistically indistinguishable from zero (at $\alpha = 0.05$ confidence level) through the alternative methodology that are thus considered `truly untrustworthy'. The revised approach results in a 98.2\% decrease in the number of points classified as untrustworthy (38 versus 2,144 with the original method), substantially increasing the amount of usable data for characterizing plasma behavior. 

Furthermore, the observations identified as untrustworthy now predominantly occur in a small window of time, latitude and longitude (further described in Figure \ref{fig:future_noise_exploration}). Although careful consideration over the entire 13 months is still needed, observations flagged as untrustworthy through the new methodology may serve as a starting point for identification of the location of the Equatorial Ionization Anomaly. This is an exciting possible direction of future research. 

\subsection*{Empirical Validation of Learned Maxima}

A fully specified generative model for raw EPEE current measurements—including all instrumental artifact mechanisms at sub-minute resolution—is not available. As a result, constructing a realistic simulation framework for validating trustworthiness classification would require strong and unverifiable assumptions about the data-generating process. Instead, we assess the reliability of the proposed peak identification procedure empirically using independent measurements from the FPMU instrument.

\begin{figure}
    \centering
    \includegraphics[width=\textwidth]{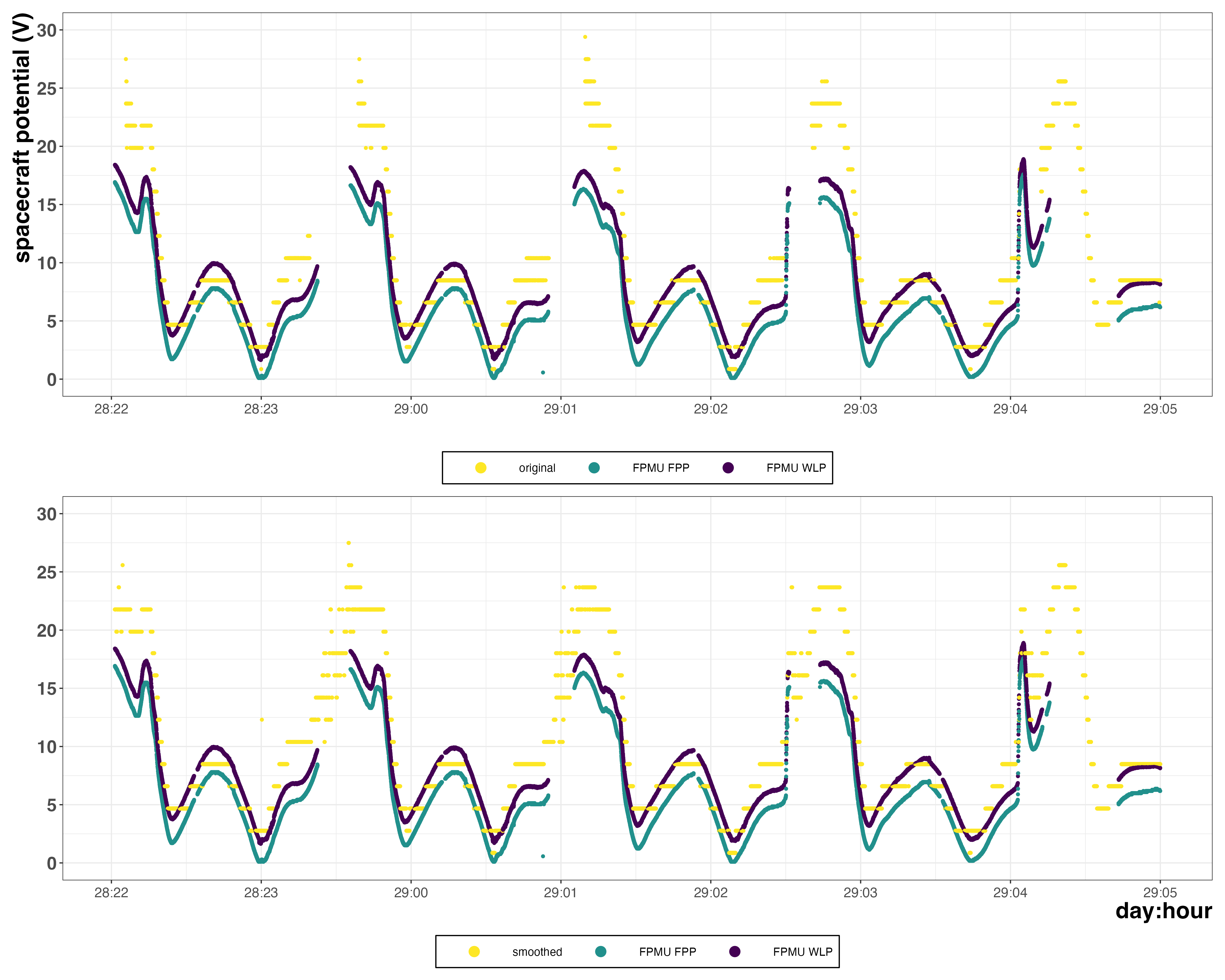}
    \caption{Spacecraft potential estimation results for both the original (top panel) and alternative (bottom panel) methods compared to estimates from the WLP (mulberry) and FPP (chartreuse) sensors on the FPMU instrument. Both original and smoothed results are in dandelion; the x-axis shows the seven hours between Julian days 271-272 shared by the two sensors in 2023.}
    \label{fig:scpotential_FPMU_calibration}
\end{figure}

Figure~\ref{fig:scpotential_FPMU_calibration} demonstrates the increased data availability for calibration with FPMU instrument data (FPP and WLP) after implementing the new methodology. In the seven-hour interval shown, the original approach produces notable gaps in usable data (top panel), limiting interpolation between calibration windows. In contrast, the smoothed current estimates (bottom panel) preserve the expected periodic structure of spacecraft charging while substantially increasing the number of usable calibration points.

Quantitatively, the cleaned EPEE currents exhibit improved coherence with FPMU
measurements, including increased correlation and reduced calibration residual variance relative to the original data. This improved cross-instrument consistency provides real-world validation that points retained by the proposed procedure reflect physically meaningful signal rather than residual artifact. The resulting increase in calibration coverage establishes a stronger foundation for estimating the more complex plasma density variable, which relies on joint interpretation of FPMU and EPEE measurements.

\section{Conclusion}\label{sec-conc}

\subsection{Potential applications}

This methodology, although specific to these data and scenario, provides a general framework applicable to other sensor interpretation tasks. Both the pipeline and the resulting smoothed energy and current surfaces can be used in the next steps of our work, which include plasma density estimation, exploration of the EIA in relation to spacecraft charging and density values, and analysis of the fluctuation of these variables in relation to the ISS orbit. These estimates also have the potential for integration to output from the International Reference Ionsophere (IRI) model. The IRI is considered the source of truth for data describing the physical parameters of the ionosphere and as of 2014 is the International Standardization Organization standard for the ionosphere \citep{Bilitza2022}. As the IRI is an empirical, data-based model, possible disadvantages arise during novel conditions. For example, during anomalous periods in the solar cycle that may not have been previously recorded, such as the very low solar cycle minimum in 2008-2009, the IRI can sometimes over- or underestimate physical parameters. In this instance, this misalignment was identified by several research groups and the IRI model was corrected \citep{Bilitza2022}. As the EPEE instrument provides data during part of Solar Cycle 25 as it heads towards the solar maximum \citep{Maldonado2023}, observations could serve to validate - and if necessary - possibly calibrate the IRI model in service to the scientific community. 

\begin{figure}
    \centering
    \includegraphics[width=\textwidth]{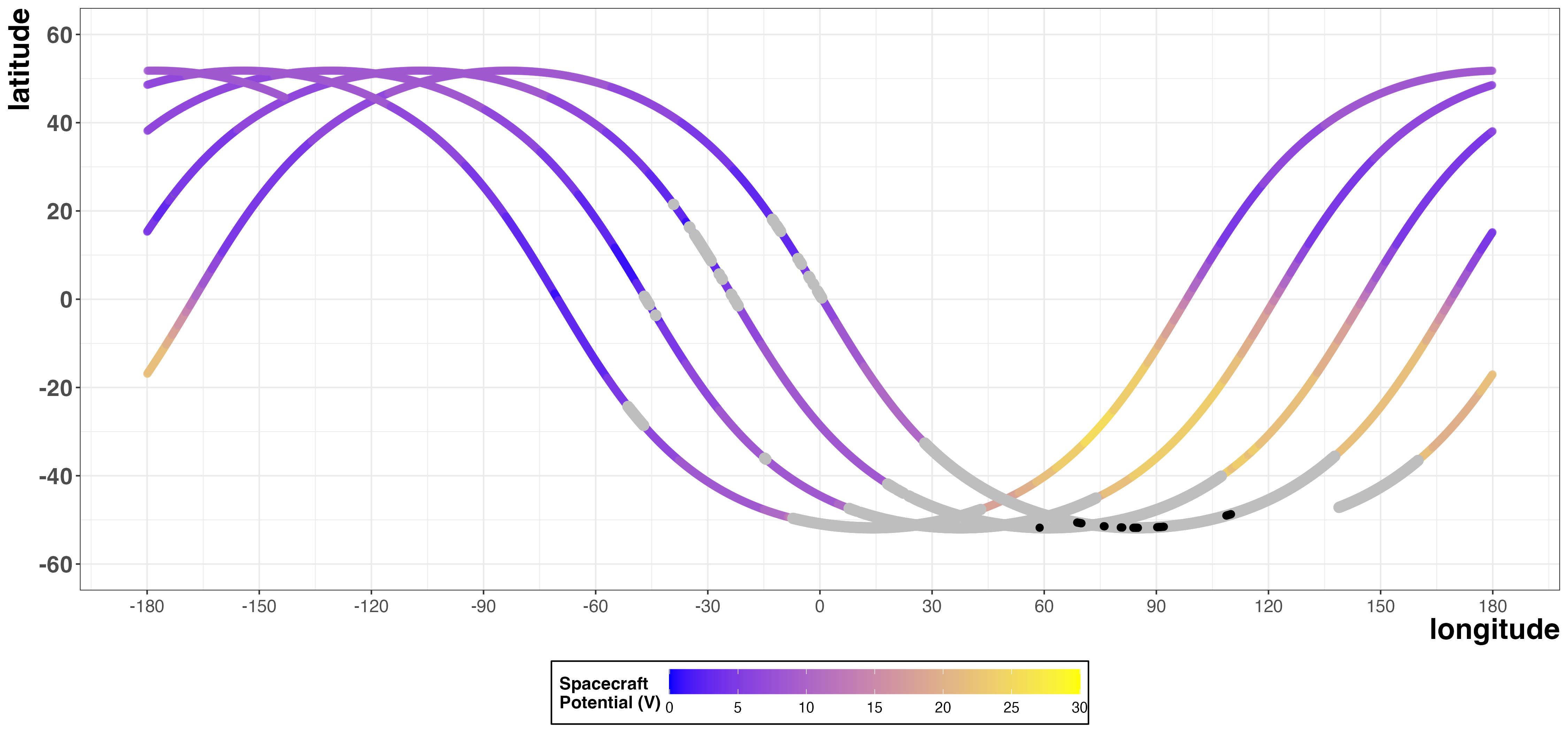}
    \caption{Spacecraft potential (V) in blue fading to yellow with increasing values are plotted across longitude (x-axis) and latitude (y-axis) for the 7-hour case study time period. The relatively small number of untrustworthy values (n=38) occur in a very concentrated latitude range and are colored in black. Previous thresholding of current values under the noise floor are in gray and show much more considerable spatial spread.}
    \label{fig:future_noise_exploration}
\end{figure}

Due in large part to this novel application of statistical methods to signal processing, we have identified the potential for trustworthiness to be used as an indicator of physical phenomena. Figure \ref{fig:future_noise_exploration} displays spacecraft potential estimates for the original data (grey) and with implementation of the new method (black) against the backdrop of latitude (y-axis) and longitude (x-axis) taken from GPS aboard the ISS. Previously, observations identified as falling below the instrument noise floor showed some scattering across locations. Now, all observations identified as untrustworthy (n=38) are concentrated within a very small latitude range, suggesting these observations may not be random. Although not pictured here, initial exploration across the entire dataset showed truly untrustworthy observations occurring in a similar spatial range. Further exploration over a longer timeline with confounding factors taken into consideration is needed. 

\subsection{Future Directions and Considerations}

Code used to process the 7-hour time period used as an example throughout this paper has been formatted to allow for batch processing across the 13-month dataset. Excitingly, our team has now implemented the non-trivial code alterations necessary to allow for heteroskedastic noise consideration within the scaled Vecchia GP approximation. Although the paper explaining these methods and subsequent ramifications is in preparation \citep{potter2025scalable}, we are ready to explore implementation for EPEE data and to compare to homoskedastic noise assumptions.

Cross calibration to the FPMU data is ongoing as part of a full data release effort by the team. This work includes a thorough review of the modified algorithms recently presented for the WLP instrument, intended "to yield
improvements in the quality of the plasma density, temperature, and potential parameters
extracted from the probe data" \citep{Minow2023}.

\section{Disclosure statement}\label{disclosure-statement}

The authors declare that no conflicts of interest exist. Research presented in this manuscript was supported by the Laboratory Directed Research and Development (LDRD) program of Los Alamos National Laboratory (LANL) under project No. 20240045DR. Work at LANL is conducted under the auspices of the United States Department of Energy. The authors also acknowledge the support of the Department of Defense (DoD) Space Test Program which provides mission design, spacecraft acquisition, integration, launch and on-orbit operations support for DoD’s science and technology (S\&T) experiments, and manages all DoD payloads on the International Space Station. Approved for public release: LA-UR-25-29600.

\section{Data Availability Statement}\label{data-availability-statement}

The data and code used in this paper have been made available at the following URL: XX. **Note: Data have been reviewed by our organization and released for public access. The code release is in progress. As soon as it is approved we can make public the git repository containing both.**

\phantomsection\label{supplementary-material}
\bigskip

\begin{center}

{\large\bf SUPPLEMENTARY MATERIAL}

\end{center}

The supplementary material contains further details describing the GP fitting, Scaled Vecchia approximation, and iterative noise profile fitting process. 

\section*{Appendix: GP and Scaled Vecchia Approximation Details}
\label{app:GP_vecchia_details}

\subsection*{Gaussian Process Parameters}

Although the GP is defined on joint coordinates $x_i=(t_i,e_i)$, we do not assume identical correlation structure in time and energy. We use an anisotropic Mat\'ern $3/2$ kernel with separate range (length-scale) parameters for time and energy, $(\ell_t,\ell_e)$, estimated from the data via maximum likelihood. In particular, correlation depends on the scaled distance
\[
r_{ij}^2 = \left(\frac{t_i - t_j}{\ell_t}\right)^2 +
\left(\frac{e_i - e_j}{\ell_e}\right)^2,
\]
allowing correlation to decay at different rates along the temporal and spectral dimensions.

\subsection{Scaled Vecchia Approximation}

The primary design parameter of the Vecchia approximation is the conditioning set size $m$, which determines the number of previously ordered neighbors in $S(i)$ used in each conditional density in \eqref{eq:Vecchapprox}. Larger values of $m$ improve approximation accuracy at increased computational cost. Following recommendations in \citet{Lawrence2022}, we fixed $m = 30$. In exploratory analyses, increasing $m$ to 40 produced no material change in parameter estimates or predicted current surfaces, indicating that $m = 30$ provided a stable approximation 
for the data considered here.

Inputs were ordered using max--min ordering with parameter-based scaling, as recommended by \citet{Lawrence2022}. Gaussian process covariance parameters (variance and range parameters of the Mat\'ern $3/2$ kernel) were estimated by maximum likelihood under the Vecchia approximation.

\section*{Appendix: Mathematical Forms of the Noise Profile Components}
\label{app:noise-model}

We model each background-dominated current profile $I_{km}(b)$ as the sum of three components: a Richards curve, a parabolic baseline correction, and a Gaussian peak. This appendix provides the explicit mathematical forms of these components and their parameterizations.

In the main text, the dependence of each model component on the energy-bin index $b$ is suppressed for clarity. Here, we provide the explicit functional forms of
$f_{km,R}$, $f_{km,P}$, and $f_{km,G}$ as functions of $b$ and their respective
parameter vectors.

\subsection*{Richards Curve}

The dominant component of the instrument baseline is modeled using a Richards curve, a generalization of the logistic function defined as
\begin{equation}
f_{km,R}(b; \boldsymbol{\theta}_R) =
A_0 +
\frac{A}{\left[1 + \nu \exp\left(-k (b - x_0)\right)\right]^{1/\nu}},
\end{equation}
where the parameter vector
\begin{equation}
\boldsymbol{\theta}_R = (A, k, x_0, \nu, A_0)
\end{equation}
consists of the amplitude at the inflection point ($A$), growth rate ($k$), inflection location ($x_0$), shape parameter ($\nu$), and vertical offset ($A_0$). This component captures the dominant monotonic structure of the instrument baseline profile.

\subsection*{Parabolic Component}

Residual low-order curvature in the instrument baseline is captured using a second-order polynomial,
\begin{equation}
f_{km,P}(b; \boldsymbol{\theta}_P) = p b^2 + q b + r,
\end{equation}
with parameter vector
\begin{equation}
\boldsymbol{\theta}_P = (p, q, r),
\end{equation}
where $p$ controls quadratic curvature, $q$ controls linear slope, and $r$ is the intercept. These parameters typically remain small in magnitude, indicating that the parabolic term provides only a minor correction to the Richards curve.

\subsection*{Gaussian Component}

Any remaining localized structure after removal of the instrument baseline is modeled as a Gaussian peak,
\begin{equation}
f_{km,G}(b; \boldsymbol{\theta}_G) =
\alpha \exp\left(
-\frac{(b - \mu)^2}{2\sigma^2}
\right),
\end{equation}
with parameter vector
\begin{equation}
\boldsymbol{\theta}_G = (\alpha, \mu, \sigma),
\end{equation}
where $\alpha$ is the amplitude of the remaining signal, $\mu$ is the peak location in energy-bin space, and $\sigma$ controls the width of the peak.

\subsection*{Combined Model}

The full fitted profile is therefore given by
\begin{equation}
I^S_{km}(b) \approx
f_{km,R}(b; \boldsymbol{\theta}_R) +
f_{km,P}(b; \boldsymbol{\theta}_P) +
f_{km,G}(b; \boldsymbol{\theta}_G),
\end{equation}
where the Richards and parabolic components together constitute the instrument baseline and the Gaussian component represents the true signal.

Figures \ref{fig:richards_parabola_fit_day272_quarter1} and \ref{fig:noiseprofile_fitestimates} both describe the selected noise profile with the smallest integral out of the candidate noise profiles for the first quarter of Julian day 272. Figure \ref{fig:richards_parabola_fit_day272_quarter1} plots the individual components that constitute the refined instrument baseline profile (Richards and parabolic curves) and identified true signal (Gaussian shape). Raw initial input and the final fitted profile (Richards and parabolic input combined) are also displayed. Figure \ref{fig:noiseprofile_fitestimates} displays the respective parameter estimates for each of these components (Richards, parabolic, Gaussian shapes) in table form. Parameters for each shape that convey the most information about the importance of the shape in the final profile fit are highlighted in colors matching the plot. These include Richards A (point of inflection), Gaussian amplitude, and the three parabolic components (quadratic term (p), slope (q), intercept (r)). 
\begin{description}
\item[Baseline decomposition with raw signal.]
\begin{center}
  \includegraphics[width=\linewidth]{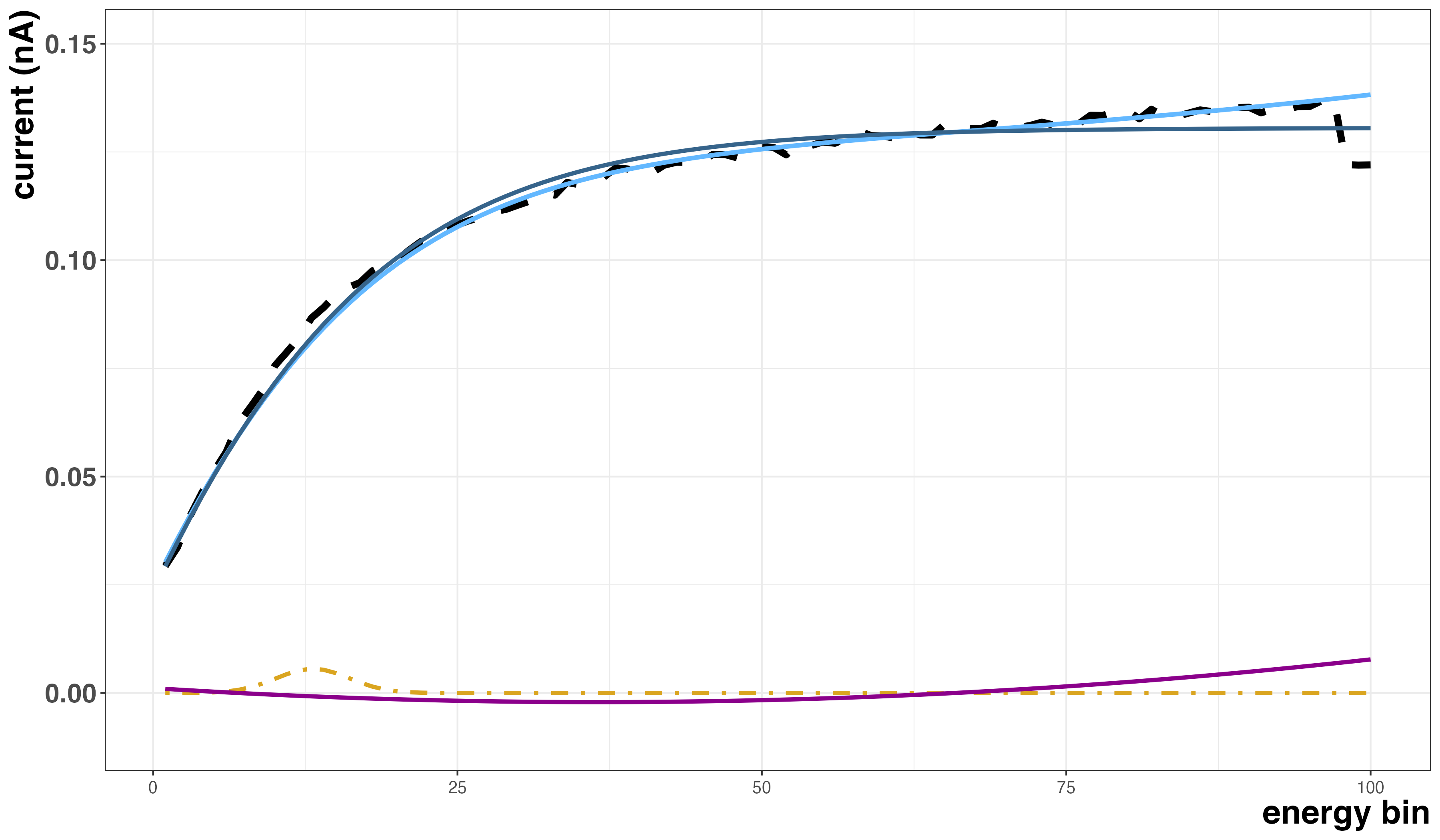}
  \captionof{figure}{Components of the noise profile fitting process for the first quarter of Julian day 272 (n=10799). The selected noise profile from all candidate noise timestamps is decomposed into Richard's fit (steel blue), parabolic (magenta) and remaining Gaussian signal (mustard). Raw signal is in dotted black and the final fitted noise profile is in cornflower blue. Energy bin is on the x-axis and current (nA) is on the y-axis.}
  \label{fig:richards_parabola_fit_day272_quarter1}
\end{center}

\item[Parameter estimates for baseline components.]
\begin{center}
  \includegraphics[width=\linewidth]{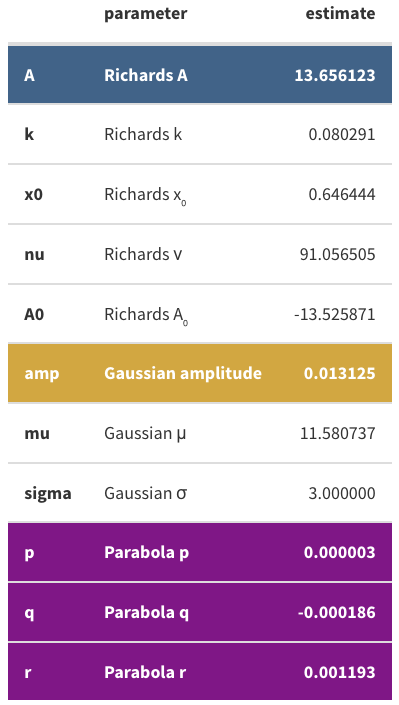}
  \captionof{figure}{Parameter estimates of the noise profile fitting process for the first quarter of Julian day 272 (n=10799) after convergence. The main parameter estimates of interest are highlighted in colors corresponding to \ref{fig:richards_parabola_fit_day272_quarter1}. Richard's A (steel blue) describes the max height at the inflection point. A relatively large value suggests that this shape dominates in the noise fitting process. The parabolic parameters (magenta) are hovering near zero, suggesting that they do not manipulate the fit much. Remaining Gaussian signal (mustard) is mainly described by amplitude, with small values indicating that little true signal remains after the Richards and parabolic iterative fitting.}
  \label{fig:noiseprofile_fitestimates}
\end{center}
\end{description}

\bibliography{bibliography.bib}

\end{document}